\documentstyle[prl,aps,twocolumn,floats,epsf]{revtex}

\newcommand{\postbb}[3]
{\setlength{\epsfxsize}{#3\hsize}
 \centerline{\epsfbox[#1]{#2}}}

\newcommand{\plb}[2]{{\em Phys. Lett.}              {\bf #1B}, #2 }
\newcommand{\npb}[2]{{\em Nucl. Phys.}              {\bf B#1}, #2 }
\newcommand{\npp}[2]{{\em Nucl. Phys. Proc. Suppl.} {\bf #1C}, #2 }
\newcommand{\pr }[2]{{\em Phys. Rep.}               {\bf  #1}, #2 }
\newcommand{\prt}[2]{{\em Phys. Rev.}               {\bf D#1}, #2 }
\newcommand{\pru}[2]{{\em Phys. Rev. Lett.}         {\bf  #1}, #2 }
\newcommand{\zpc}[2]{{\em Z. Phys.}                 {\bf C#1}, #2 }

\newcommand{\app}[2]{{\em Acta Phys. Polon.}        {\bf B#1}, #2 }

\newcommand{\ijmpa}[2]{{\em Int. J. Mod. Phys.}     {\bf A#1}, #2 }

\newcommand{\jpl}[2]{{\em JETP Lett.}               {\bf  #1}, #2 }
\newcommand{\epc}[2]{{\em Eur. Phys. J.}            {\bf C#1}, #2 }
\newcommand{\con}[2]{                               {\bf  #1}, #2 }
\newcommand{\etal}{{\em et al.}}
\newcommand{\ibid}{{\em ibid.}}

\newcommand{\as}{\hat\alpha_s}
\newcommand{\aspi}{{\hat\alpha_s\over \pi}}
\newcommand{\be}{\begin{equation}}
\newcommand{\ee}{\end{equation}}
\newcommand{\ba}{\begin{array}}
\newcommand{\ea}{\end{array}}

\def\ms{\mbox{{\footnotesize{$\overline{\rm MS}$}}} }

\newcommand{\lsim}{\buildrel < \over {_\sim}}
\newcommand{\gsim}{\buildrel > \over {_\sim}}

\begin{document}

\draft

\title{\hfill {\normalsize UPR--796--T} \\ \hfill \\
Calculation of the QED Coupling $\hat\alpha (M_Z)$ in the Modified
Minimal-Subtraction Scheme}
\author{Jens Erler}
\address{Department of Physics and Astronomy, University of Pennsylvania, 
Philadelphia, PA 19104-6396, USA}

\date{March 1998}

\maketitle

\begin{abstract}
I calculate the QED coupling, $\hat\alpha$, directly in the \ms scheme 
using an unsubtracted dispersion relation for the three light quarks, 
and perturbative QCD for charm and bottom quarks. Compact analytical 
expressions are presented, making this approach particularly suitable for 
electroweak fits. After $\hat\alpha^{-1} (m_\tau) = 133.513 \pm 0.026$ 
is obtained in a first step, I perform a 4-loop renormalization group 
evolution with 3-loop matching conditions to arrive at 
$\hat\alpha^{-1} (M_Z) = 127.934 \pm 0.027$ for $\hat\alpha_s (M_Z)= 0.120$. 
The corresponding hadronic contribution to the on-shell coupling is 
$\Delta\bar\alpha_{\rm had}^{(5)} (M_Z) = 0.02779 \pm 0.00020$. The error 
is mainly from $m_c$, and from experimental uncertainties in $e^+ e^-$ 
annihilation into unflavored and strange hadrons and $\tau$ decay data.
\end{abstract} 


\section{Intorduction}

The increasing precision of experiments at CERN, SLAC, and the Tevatron
calls for refined theoretical calculations with corresponding accuracy. 
For example, the precision of measurements of $M_W$, the 
$Z$ width, and the weak mixing angle, $\hat{s}^2_Z$, has reached or surpassed 
the per mille level. Most theoretical uncertainties are presently still 
negligible compared to the experimental ones. In contrast, the QED coupling 
constant (\ms quantities will be marked by a caret),
\be
\label{alphahat}
   \hat\alpha(\mu) = {\alpha\over 1 - 4\pi\alpha\hat\Pi(0)}, 
\ee
escapes a precise theoretical computation from the fine structure constant, 
$\alpha$, for $\mu > 2 m_{\pi^0}$ when hadronic effects must be included. 
On the other hand, knowledge of $\hat\alpha (M_Z)$ is indispensable for the 
extraction of the Higgs mass from precision data. In particular, with 
$m_t$ known independently from top quark production at the Tevatron, $M_W$ and 
$\hat{s}^2_Z$ now serve as the most important Higgs probes, but are strongly 
correlated with $\hat\alpha (M_Z)$. Clearly, precise and reliable information 
on $\hat\alpha (M_Z)$ is needed. 

This has prompted a great deal of activity in the course of the past 
4 years~\cite{Nevzorov94,Martin95,Davier98,Kuhn98,Alemany98}. The traditional 
strategy is to exploit the analytic properties of QCD and obtain a subtracted 
dispersion relation (SDR)~\cite{Kniehl96} for the on-shell coupling,
$\bar\alpha (M_Z)$. By virtue of the optical theorem a weighted integral over 
the $e^+ e^-$ cross section ratio, 
\be
   R(s)=12\pi {\rm Im\,}\hat\Pi^{\rm (had)}(s) = {\sigma_{\rm hadrons}\over
   \sigma_{\mu^+ \mu^-}},
\ee 
is obtained. The task is now reduced to construct the function
$R(s)$ for the entire energy regime $s \geq 4 m_{\pi^0}^2$
using both theoretical and experimental information. This is a complex process,
and the various papers~\cite{Nevzorov94,Martin95,Davier98,Kuhn98,Alemany98}
differ by the methods by which experimental data are averaged, by the energy 
regimes in which perturbative QCD (PQCD) is trusted, by the treatment and 
parametrization of resonances, etc. The situation is aggravated by the fact 
that the data sets in question often lack a thorough documentation, and much of
the data taking was done at a time when per mille precisions were not 
anticipated. Consequently, an emancipation from old or imprecise data is 
indicated whenever possible. 

The key question is the energy domain in which the use of PQCD is adequate. 
PQCD and QCD driven data normalization for energies as low as a few GeV have 
been advocated by Martin and Zeppenfeld~\cite{Martin95}, but at the time 
encountered scepticism. A breakthrough came from three sides. First, 
theoretical calculations~\cite{Braaten92} and measurements of the $\tau$ 
lifetime have matured significantly over the last decade, with the extracted 
strong coupling constant~\cite{Hocker97}, $\hat\alpha_s$, in perfect agreement 
with the value extracted from the $Z$ lineshape~\cite{Erler98}. 
Secondly, in Ref.~\cite{Hocker97} H\"ocker studies the invariant mass 
distribution in $\tau$ decays following the spectral moments proposed by 
Le Diberder and Pich~\cite{LeDiberder92}. When varying his input data, he 
obtains impressively consistent results for the non-perturbative contributions 
to vector and axial-vector two-point correlation functions, which are moreover 
fairly small. Having passed this test, Davier and H\"ocker~\cite{Davier98} 
finally fit to analogously defined spectral moments of $R(s)$, with the 
conclusion that non-perturbative contributions to $R(m_\tau)$ are negligibly 
small. Motivated by these developments, K\"uhn and Steinhauser~\cite{Kuhn98} 
present a state of the art analysis of the SDR approach.

In this letter I introduce a different method based on an unsubtracted 
dispersion relation (UDR). This is the natural framework for the computation 
of $\hat\alpha (\mu)$, defined at only one momentum transfer, $q^2 = 0$. 
By its very concept it is a purely perturbative quantity. Complications from 
non-perturbative physics arise only to the extent to which dispersion 
relations are used. This is hard to avoid for the three light flavors, 
but $c$ and $b$ quarks will turn out to be massive enough to be treated 
exclusively within PQCD. The approach chosen in this paper also has an 
important practical advantage. Calculations of Higgs limits or 
$\chi^2$ plots~\cite{Erler98} require thousands of fits each with multiple 
function calls. A numerical (dispersion) integration within each call would be 
too expensive computationally, but within the approach introduced in this work,
no numerical integration will be necessary. As a result $\hat\alpha (M_Z)$ can 
be self-consistently recalculated in each call, and the parametric uncertainty 
due to $\hat\alpha_s$ (which is a fit parameter) can be dropped. 

\section{Heavy Quarks}

The polarization function in Eq.~(\ref{alphahat}) is defined through the 
current correlator,
$$
   (q_\mu q _\nu - q^2 g_{\mu\nu}) \hat\Pi (q^2) = 
   i \int d^4 x e^{iqx} \langle 0 | T j_\mu (x) j_\nu (0) | 0 \rangle,
$$
where $j_\mu$ is the electromagnetic current. For a heavy quark it has been 
calculated up to 3-loop ${\cal O} (\hat\alpha\hat\alpha_s^2)$ in 
Ref.~\cite{Chetyrkin96}. The result for $\hat\Pi^{(h)} (0)$ is expressed in 
terms of the quark pole mass. The coefficients grow rapidly and application 
to charm (bottom) quarks is impossible (questionable). However, the adverse 
coefficients are almost entirely due to the employment of the pole mass, 
which is (due to quark confinement) not a well defined 
quantity~\cite{Chetyrkin96A}. It is therefore appropriate to reexpress 
$\hat\Pi^{(h)} (0)$ in terms of the \ms mass, $\hat{m} (\mu)$, yielding
(in agreement with Ref.~\cite{Chetyrkin98}),
$$
   \hat\Pi^{(h)} (0) = {Q_h^2\over 4 \pi^2} \left\{L + 
   \aspi \left[{13\over 12} - L \right]
   + {\as^2\over \pi^2} \left[{655\over 144}\zeta(3) - \right.\right.
$$
\be
\label{pi0h} 
   \left.\left. {3847\over 864} - {5\over 6} L - {11\over 8} L^2 + n_q \left(
   {361\over 1296} - {L\over 18} + {L^2\over 12}\right) \right] \right\},
\ee
where $Q_h$ is the electric charge of the heavy quark, $n_q$ the number of 
active flavors, and $L = \ln {\mu^2\over \hat{m}^2}$. Now all coefficients are 
of order unity, indicating a reliable expansion. Moreover, all terms 
proportional to $\pi^2$ have cancelled. By setting $\mu = \hat{m}(\mu)$ the 
$L$ terms can also be dropped. The remaining constant terms play the r\^{o}le 
of matching coefficients to be applied when the number of flavors in the 
effective theory is increased from $n_q - 1$ to $n_q$. This is familiar from 
the renormalization group evolution (RGE) of $\as$. Since $n$-loop matching
must be supplemented with $n + 1$-loop RGE, inclusion of the 
${\cal O} (\alpha \as^3)$ beta function contribution is required and will be 
discussed later.

Eq.~(\ref{pi0h}) describes the contribution of a heavy quark in the external 
current. The $n_q - 1$ light quarks appearing in internal loops must be 
treated as massless, since 3-loop diagrams involving two massive quarks with 
different masses have not been computed. It is indeed safe to neglect terms of 
${\cal O} (\as^2 \hat{m}_l^2/\hat{m}_h^2)$, since in practice the heavy quark 
mass, $\hat{m}_h$, is always sufficiently larger than all lighter quark masses,
$\hat{m}_l$, and we will follow this approximation throughout. Conversely, in 
${\cal O} (\as^2)$ the heavy quark also appears as a loop insertion into a 
one-gluon exchange diagram (the ``double bubble'' diagram), and in the wave 
function renormalization of a light quark in the external current. 
The limit $q^2 \rightarrow 0$ can only be performed when the heavy quark is 
decoupled, i.e., the $\as$ definition for $n_q - 1$ active flavors is used. 
This has been done in Ref.~\cite{Chetyrkin98},
\be
\label{pi0gs} \hspace{-2pt}
   \delta\hat\Pi^{(h)} (0) = \sum\limits_l {Q_l^2\over 4\pi^2} 
   {\as^2\over \pi^2} \left({295\over 1296} - 
   {11\over 72} L + {L^2\over 12} \right). \hspace{-2pt}
\ee
Eqs.~(\ref{pi0h}) and (\ref{pi0gs}) carry to ${\cal O} (\as^2)$ the decoupling
of a heavy quark~\cite{Chetyrkin98}, as was first suggested by Marciano and
Rosner~\cite{Marciano90} for the case of the top quark and generalized to 
${\cal O} (\as)$ in Ref.~\cite{Fanchiotti93}. The same decoupling can also
be applied to the \ms definition of the weak mixing angle.

With $\as (\hat{m}_c)/\pi \approx 0.13$ and the absence of non-perturbative
effects, Eqs.~(\ref{pi0h}) and (\ref{pi0gs}) can be used reliably not only 
for $b$ but also for $c$ quarks. Complications with $J/\Psi$ and $\Upsilon$ 
resonances are then completely avoided at the expense of the introduction of 
a stronger dependence on the quark masses. The numerical uncertainty due to 
$\hat{m}_b$ will turn out to be small, while $\hat{m}_c$ will introduce an 
error comparable to the one introduced through the $J/\Psi$ resonances in the 
SDR approach. 

\section{Light Quarks}

\begin{figure}[t]
\postbb{-30 -30 430 200}{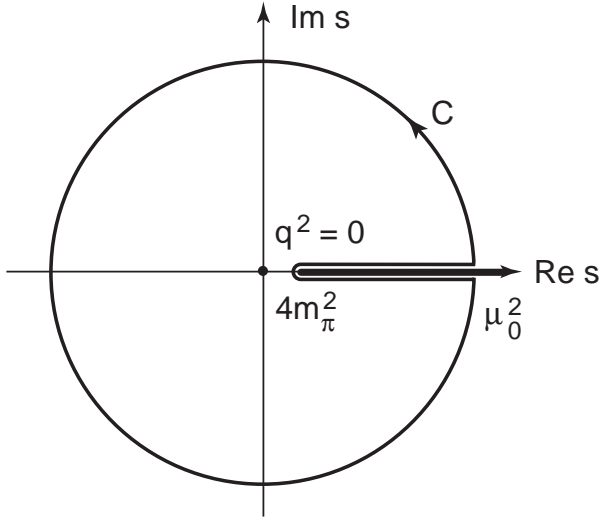}{1.8}
\caption{ Contour for an unsubtracted dispersion integration.}
\label{fig}
\end{figure}

I now turn to the three light quark flavors. Applying Cauchy's theorem to
the contour in Fig.~\ref{fig}, yields
\be
   \hat\Pi(0) = {1\over \pi} \int\limits_{4 m_\pi^2}^{\mu_0^2}
   {ds\over s - i\epsilon}\, {\rm Im\,} \hat\Pi(s) + {1\over 2\pi i} 
   \oint\limits_{|s| = \mu_0^2} {ds\over s} \hat\Pi(s).
\ee
The optical theorem applied to the first term, and the substitution 
$s = \mu_0^2 e^{i\theta}$ to the second, brings the UDR into its final form,
\be
\label{UDR}
   \hat\Pi^{(3)}(0) = {1\over 12\pi^2} \int\limits_{4 m_\pi^2}^{\mu_0^2} 
   {ds\over s - i\epsilon} R(s) + {1\over 2\pi} \int\limits_{0}^{2\pi} 
   d\theta\, \hat\Pi^{(3)}(\theta).
\ee
As in the SDR approach, the first integral can be evaluated using the measured
function $R(s)$ up to a scale $\mu_0$ where PQCD is trusted. Together with 
the second (called $I^{(3)}$ hereafter) this results for $\mu_0 < M_{J/\Psi}$ 
in the 3-flavor definition $\hat\alpha^{(3)} (\mu_0)$. Other values of $\mu$ are
obtained using RGE, and other quark and lepton flavors are included at 
$\mu = \hat{m} (\mu)$ using the matching description discussed before. Special 
care is needed if $\mu_0 > \hat{m}_c$, where conventionally 4-flavor QCD is 
used. The clash with 3-flavor QED will generate some extra (non-decoupling) 
logarithms. Indeed, following Refs.~\cite{Davier98,Kuhn98} I will use 
$\mu_0 = 1.8$~GeV and the result of Ref.~\cite{Alemany98}, as quoted in 
Ref.~\cite{Davier98},
\be
\label{SDR}
   {\alpha M_Z^2\over 3\pi} \int\limits_{4 m_\pi^2}^{\mu_0^2} 
   ds {R(s)\over s(M_Z^2 - s) - i\epsilon} = (56.9 \pm 1.1) \times 10^{-4}.
\ee
The difference between this and the first integral in Eq.~(\ref{UDR}) 
(times $4\pi\hat\alpha$) can be neglected since,
$$
   {\alpha\over 3\pi} \int\limits_{4 m_\pi^2}^{\mu_0^2} ds R(s) \left[
   {1\over s - i\epsilon} - {M_Z^2\over s(M_Z^2 - s) - i\epsilon} \right] =
$$
\vspace{-15pt}
$$
   2\alpha \mu_0^2 \int\limits_{0}^{2\pi} d\theta {\hat\Pi^{(3)}
   (\mu_0^2 e^{i\theta}) \over M_Z^2 e^{- i\theta} - \mu_0^2} 
   \approx - {2\alpha\mu^2_0\over 3\pi M_Z^2} \approx - 6 \times 10^{-7}.
$$
The second integral in 
Eq.~(\ref{UDR}) can again be obtained with the help of Ref.~\cite{Chetyrkin96},
\be
\label{int}
   I^{(3)} = {1\over 6\pi^2} \left\{ {5\over 3} + \aspi 
   \left[{55\over 12} - 4\zeta(3) + 2 {\hat{m}_s^2 (\mu_0)\over \mu_0^2}
   \right] + \right.  
\ee
\vspace{-15pt}
$$
   \left. {\as^2\over \pi^2} \left[{34525\over 864} - {9\over 4}\zeta(2) - 
   {715\over 18}\zeta(3) + {25\over 3}\zeta(5) + 
   F({\mu_0^2\over \hat{m}_c^2}) \right] \right\},
$$
where in the ${\cal O} (\as)$ term I kept the small $s$ quark mass effect 
($\sim 2\times 10^{-6}$). $F(x)$ can be reconstructed from the absorbtive 
part of $\hat\Pi^{(3)} (s)$, i.e., $R(s)$. Below threshold it can be well 
approximated as an expansion in $x$, despite the fact that 
$\hat{m}_c^2 < \mu_0^2$, as can be shown by comparison with the exact 
result~\cite{Kniehl90} (the large quark mass expansion in ${\cal O} (\as^3)$
is also known~\cite{Larin95}). The coefficients in $F(x)$ decrease even more 
rapidly than in $R(s)$,
$$
  F(x) \approx \ln x \left[{2\over 3}\zeta(3) - {11\over 12} + {\ln x\over 12}
  \right] - x \left[{2 \over 25} - {2 \over 135} \ln x \right] 
$$
\vspace{-15pt}
$$
  + x^2 \left[{1513\over  2116800} - {\ln x\over 5040}   \right]
  - x^3 \left[{1853\over 80372250} - {\ln x\over 127575} \right].
$$
In principle, $F(x)$ also applies to the tiny $b$ quark contribution to 
$\hat\Pi^{(3)} (0)$, but without the non-decoupling logarithms in the first 
term.

\section{Renormalization Group Evolution}

The QED $\beta$ function including QCD corrections reads,
$$
   \beta \equiv \mu^2 \frac{d\hat\alpha}{d\mu^2} =
   4\pi \hat\alpha^2 \mu^2 \frac{d\, \hat\Pi(0)}{d\mu^2} \equiv
- {\hat\alpha^2\over \pi} (
       \beta_0 + 
       \beta_1 {\hat\alpha\over \pi} +
   \hat\beta_2 {\hat\alpha^2\over \pi^2}  
$$
\vspace{-15pt}
\be
\label{RGE} 
    + \delta_1 \aspi +
   \hat\delta_2 {\hat\alpha_s^2\over \pi^2} + 
   \hat\delta_3 {\hat\alpha_s^3\over \pi^3} + 
   \hat\delta_4 {\hat\alpha_s^4\over \pi^4} +
       \epsilon_2{\hat\alpha\over \pi} \aspi + \dots),
\ee
where coefficients with a caret are scheme dependent.
$$ 
   \beta_0 = -         \sum\limits_{f} {Q_f^2\over 3}, \;\;
   \beta_1 = -         \sum\limits_{f} {Q_f^4\over 4}, \;\;
  \delta_1 = - N_c C_F \sum\limits_{q} {Q_q^2\over 4},
$$ 
and $\epsilon_2$ are scheme independent and can be gleaned from the 3-loop 
$\beta$ function for simple groups~\cite{Tarasov80}. $\hat\delta_2$ is 
straightforwardly computed from Eq.~(\ref{pi0h}), resulting in
\be
\label{delta2}
   \hat\delta_2 = N_c C_F \sum\limits_{q} Q_q^2 \left[{1\over 32} C_F -
                  {133\over 576} C_A + {11\over 144} T_F n_q \right],
\ee
where for QCD we have 
$N_c = C_A = 3$, $C_F = 4/3$, and $T_F = 1/2$. $\epsilon_2$ 
can be obtained from the first term with the substitution 
$Q_q^2 C_F \rightarrow Q_q^4$. The coefficients in the first and the last 
term of Eq.~(\ref{delta2}) are familiar from
$$
  \hat\beta_2 = {1\over 32} \sum\limits_{f} Q_f^6 + {11\over 144} 
  \left(\sum\limits_{f} Q_f^4\right) \left(\sum\limits_{f} Q_f^2\right).
$$
On the other hand, the second term cannot be obtained from 
Refs.~\cite{Tarasov80}, since it cannot be disentangled from contributions 
with gluons in the external current. $\hat\delta_3$ has been obtained by
Chetyrkin~\cite{Chetyrkin97}, and can also be reconstructed in the following 
way. Analytical continuation encodes the 4-loop order logarithms of 
$\hat\Pi (0)$ in
$$R(s) = N_c \sum_{q} Q_q^2 \sum_{i} r_i {\hat\alpha_s^i \over \pi^i},$$
where the $r_i$ are the non-singlet coefficients (singlet contributions which 
are to be treated likewise are ignored for the moment). Denoting QCD $\beta$ 
function coefficients by $\beta_i^{(3)}$, and the constant terms appearing in 
Eq.~(\ref{int}) in two and three-loop order by 
$\rho_2 = C_F (55/16 - 3 \zeta(3))$ and $\rho_3$, I find from a comparative
analysis of leading logarithms in the SDR and UDR approaches,
\be
{\hat\delta_2\over \delta_1} = r_2 -\beta_0^{(3)}\rho_2, \hspace{20pt}
{\hat\delta_3\over \delta_1} = r_3 -\beta_1^{(3)}\rho_2 - 2\beta_0^{(3)}\rho_3.
\ee
With the $r_i$ 
from Refs.~\cite{Larin95,Gorishny88} I confirm Eq.~(\ref{delta2}), and find,
$$
   \hat\delta_3 = N_c C_F \sum\limits_{q} Q_q^2 \left[ {23\over 128} C_F^2 
       - \left({215\over 864} - {11\over 72} \zeta(3) \right) C_F C_A \right.
$$
\vspace{-10pt}
$$
       - \left({5815\over 62208} + {11\over 72} \zeta(3)\right) C_A^2
       + \left({169\over 864} - {11\over 36} \zeta(3) \right) C_F T_F n_q
$$
\vspace{-10pt}
$$ \left.
       + \left({769\over 15552} + {11\over 36} \zeta(3)\right) C_A T_F n_q
       + {77\over 3888} T_F^2 n_q^2 \right] 
$$
\vspace{-10pt}
\be
\label{delta3} 
       - \left(\sum\limits_{q} Q_q\right)^2
         \left({11\over 144} - {1\over 6} \zeta(3)\right) T_F^2 d^{abc}d_{abc},
\ee
where for QCD $T_F^2 d^{abc}d_{abc} = 10/3$. Eqs.~(\ref{delta2}) 
and~(\ref{delta3}) agree with Ref.~\cite{Chetyrkin97}. Some of the terms in 
Eq.~(\ref{delta3}) can also be checked with 
the four-loop QCD $\beta$ function~\cite{vanRitbergen97}. There are delicate
cancellations for $n_q = 4$ and 5, for which $\hat\delta_3 = -1.21$ and $1.23$,
respectively (for 6 quarks $\hat\delta_3 = 5.79$), and the next order might be
larger without this being an indication of a breakdown of perturbation theory. 
By assuming no cancellations in $\hat\delta_4$, a conservative estimate of 
higher order RGE contributions is $|\hat\delta_4/\delta_1| \leq C_A^3$. With
$$
   c_i = {\beta_i^{(3)}\over \beta_0^{(3)}}, \;\;\;\;
   a_0 = {\hat\alpha_s (\mu_0)\over \pi},   \;\;\;\;
   L = \ln {\mu^2\over \mu_0^2},    \;\;\;\;
   X = a_0 \beta_0^{(3)} L,
$$
and the approximation,
\be
   Y = \ln (1 + X) \lsim {\cal O} (1),
\ee
Eq.~(\ref{RGE}) can be solved 
analytically, which is welcome for numerical implementations:
$$ 
   {\pi\over \hat\alpha (\mu)} - {\pi\over \hat\alpha (\mu_0)} = \beta_0 L + 
   {\beta_1\over \beta_0} \ln (1 + {\hat\alpha (\mu_0)\over \pi}\beta_0 L) +
   {\delta_1\over \beta_0^{(3)}} Y + 
$$
\vspace{-10pt}
$$
   {1\over \beta_0^{(3)}}{a_0\over 1+X}[\hat\delta_2 X + \delta_1 c_1(Y - X)]+
   {1\over \beta_0^{(3)}} \left( {a_0\over 1 + X} \right)^2 \times
$$
\vspace{-10pt}
$$   
   [(\hat\delta_3 - \hat\delta_2 c_1) (X + {X^2\over 2}) + 
    \delta_1(c_1^2 - c_2) {X^2\over 2} + 
     \hat\delta_2 c_1 Y - \delta_1 c_1^2 {Y^2\over 2}].
$$
Effects from $\hat\beta_2\approx 2$, and $\epsilon_2 \approx 0.05$, are well 
below $10^{-6}$, and can be ignored. QED corrections to the QCD $\beta$
function induce a contribution which is formally of the same order as 
$\epsilon_2$. It turns out to be negligible, as well.

\section{Other Contributions}

To complete the calculation of $\hat\alpha (\mu_0)$, non-hadronic contributions
have to be added to Eq.~(\ref{UDR}),
$$
   \hat\Pi^{\rm (non-had)} (0) = {1\over 4\pi^2} \left\{ 
   \left( {1\over 3} + {\hat\alpha\over 4\pi} \right)
   \left( \ln {\mu_0^2\over m_e^2} + \ln {\mu_0^2\over m_\mu^2} \right)\right.
$$
\be
\label{lep} \left.
  + {\hat\alpha^2\over 24 \pi^2} \left( \ln^2 {\mu_0^2\over m_e^2} 
  + 3\ln^2 {\mu_0^2\over m_\mu^2}\right) + {15\over 8} - {1\over 6}\right\}.
\ee
The logarithms are the RGE effects of electrons and muons up to ${\cal O}
(\hat\alpha^3)$. For consistency, only the leading logarithms should be 
included in 3-loop order. 
Leptonic ${\cal O} (\hat\alpha^3)$ results are also available in the 
literature~\cite{Steinhauser98}. The third term is the corresponding matching 
effect analogous to the $13/12$ term in Eq.~(\ref{pi0h}). At this point I 
should stress that \ms masses are used only as far as QCD is concerned; when 
quark effects (or leptons) get QED corrected, the mass is treated as on-shell.
The last term is the bosonic contribution~\cite{Fanchiotti93,Sirlin80} 
(the $W^\pm$ contribution to $\beta_0$ is $+7/4$). ${\cal O} (\hat\alpha^2)$ 
correction are included for fermions, but not for bosons, because 2-loop 
electroweak calculations are generally unavailable. 

\section{Non-perturbative Effects}

Thus far the discussion has been entirely within perturbation theory. 
As stated in the introduction, non-perturbative contributions are expected 
to be small. In this section I will discuss the uncertainties associated with 
possible non-perturbative effects.

The operator product expansion (OPE)~\cite{Wilson69,Shifman79} supplements 
perturbation theory with terms suppressed by powers of~$s$. Dimension~2 terms 
can only arise from an expansion in the light quark masses. Therefore, $D=2$ 
operators can be treated perturbatively and the strange quark mass effect is 
already included in Eq.~(\ref{int}). Dynamical operators appear only at $D=4$ 
and higher. For example, the strange quark and gluon 
condensates~\cite{Shifman79} are the dynamical operators of $D=4$ and give 
rise to the contributions,
\be
\label{condensates}
   \Delta I^{(3)} = {1\over 6\pi^2} {\as^2\over \pi^2} \left[
   {7\pi^2\over 6} {\langle m_s \bar{s} s \rangle \over \mu_0^4} - 
   {11\pi^2\over 48} {\langle {\alpha_s\over \pi} GG \rangle \over \mu_0^4}
   \right],
\ee
where the condensates are of order $- m_K^2 f_\pi^2$ and $\Lambda_{\rm QCD}^4$,
respectively. Note, that these terms are suppressed by two powers of 
$\hat\alpha_s$ and therefore very small. They change $\hat\alpha (m_\tau)$ by 
about $- 2 \times 10^{-7}$, an effect completely negligible. Effects from up 
and down quark condensates are suppressed by a further factor of 
$m_\pi^2/m_K^2$, and quartic mass terms are tiny, as well. 

As for the $c$ and $b$ quark contributions, Eq.~(\ref{pi0h}) orginates from 
a heavy quark expansion and there is no quark vacuum expectation value. 
There is, however, a small contribution from the gluon 
condensate~\cite{Shifman79,Broadhurst94},
\be
\label{hcondensate}
   \Delta \hat\Pi^{(h)} (0) = {Q_h^2\over 4 \pi^2} \left[
   - {\pi^2\over 30} \left(1 + {605\over 162} \aspi \right)
   {\langle {\alpha_s\over \pi} GG \rangle \over \hat{m}^4} \right],
\ee
which is negligible for $b$ quarks, and is between $-3$ and 
$- 7 \times 10^{-6}$ for $c$ quarks, depending on the employed value
for the condensate. This is still below other non-perturbative uncertainties
discussed in the following. In conclusion, non-perturbative effects which 
can be described by local operators within the OPE are well under control. 
I will argue below, that this fact can also be used to limit possible other 
non-perturbative contributions. 

While the OPE takes phenomenologically into account a class of 
non-perturbative effects, its coefficient functions are still computed by 
perturbative means. Therefore, it cannot fully assess truly non-perturbative 
effects proportional to $\exp( - {{\rm const}/\alpha_s})$. To discuss the 
validity of the OPE requires an understanding of such effects. The leading
correction to the perturbative treatment at short distances is believed to be
associated with the one-instanton solution~\cite{Shifman79,tHooft76}. 
In pure\footnote{The inclusion of (almost) massless quarks rather extends the 
validity of the OPE~\cite{Shifman79}.} QCD the density for instantons of small 
size, $\rho$, is proportional to~\cite{tHooft76,Callan78}
\be
\label{instantons}
   {{\rm d} \rho \over \rho^{2n+1}} \left({2\pi\over \alpha_s}\right)^6 
   e^{ - {2\pi\over \alpha_s}},
\ee
where the integer $n\geq 2$ can be fixed on dimensional grounds. Clearly, this
density does not apply to large size instantons with 
$\rho\gsim \Lambda_{\rm QCD}^{-1}$, which contribute unsuppressed. For example,
one expects a large size instanton contribution of order $\Lambda_{\rm QCD}^4$
to the gluon condensate ($n=2$). We lack a full understanding of instanton 
effects, but large size instanton contributions are described with the 
phenomenological parameters of the OPE. 

When the expression~(\ref{instantons}) is integrated over $\rho$, one would 
encounter ultraviolet singularities for large enough $n$. This can be seen by 
changing the integration variable first to $\mu = \rho^{-1}$, and then to 
\be
  \alpha_s (\mu) = {\pi\over \beta_0^{(3)} \ln (\mu^2/\Lambda_{QCD}^2)}.
\ee
Inserting the lowest order $\beta$-function coefficient for pure QCD 
would result in ultraviolet divergences for $n \geq 6$~\cite{Shifman79} (see
also Eq.~(\ref{alpha_center}) below). The interpretation of these divergences 
is that the integral is actually cut off at $\mu \sim \mu_0$, where $\mu_0$ is
the energy scale of the problem at hand. This limits the applicability of 
the OPE in two ways: (1) It introduces scale dependences in the matrix 
elements which are in conflict with the separation of short and large distance
dynamics within the OPE. (2) Starting from some larger value of $n$ there will
be no suppression of higher order power terms. 

In order to be able to trust results derived within the OPE, one should 
therefore try to limit possible small instanton contributions. The small size 
instanton density~(\ref{instantons}) can be integrated for small $n$,
\be
\label{instantons2}
   {A_n\over 720} \int\limits_{0}^{\infty} {{\rm d}\alpha_s\over\alpha_s}
   \rho^{-2n} \left( {2\pi\over\alpha_s} \right)^7 e^{-{2\pi\over\alpha_s}},
\ee
where the $A_n$ are parameters of ${\cal O} (1)$ ($A_0 = 1$ by normalization).
The $\alpha_s$ distribution is centered around 
$\alpha_s = {\pi\over 3} (1\pm {1\over \sqrt{5}})$. 
For $\alpha_s(M_Z) = 0.120$ this would correspond to instanton sizes 
$\rho\sim 1.5$ GeV$^{-1}$. However, the extra suppression factor 
$\rho^{-2n} \sim \mu^{2n}$ effectively shifts the distribution center 
to smaller values:
\be
\label{alpha_center}
   \alpha_s = {\pi\over 3} (1 - {n\over 2 \beta_0^{(3)}}) 
   (1\pm {1\over \sqrt{5}}).
\ee
For large enough $n$ the small instanton contribution will be dominated by the 
energy scale $\mu_0$. One can thus read off the suppression factor for OPE
breaking effects,
\be
\label{suppression}
   {A\over 720\alpha_s} \left( {2\pi\over \alpha_s} \right)^7
   e^{-{2\pi\over\alpha_s}} \sim 0.03 A,
\ee
where $A$ is a collective parameter again of ${\cal O} (1)$, and
$\alpha_s = \alpha_s (\mu_0)$.

While one cannot estimate the parameters $A_n$, one can try to put an order
of magnitude bound on $A_2$ using the phenomenological value of the two gluon
condensate~\cite{Davier98},
\be
\label{condensate}
    \langle {\alpha_s\over \pi} GG \rangle \approx 0.04 \mbox{ GeV}^4.
\ee
I will assume that it is dominated by small rather than large instantons. 
This assumption will allow an order of magnitude bound on small instanton 
effects if one ignores the possibility of large cancellations. For $n=2$ 
the integral~(\ref{instantons2}) is contributed on average by instantons of 
scale $\mu_I\sim 0.9$~GeV. Using that one can conclude, 
$A_2 \lsim 0.07$.

Even assuming that small instantons are (unlike the OPE power terms appearing 
in $I^{(3)}$) not suppressed by further powers of $\hat\alpha_s/\pi$, one 
would find from Eq.~(\ref{suppression}) with $A=1$ a suppression factor of 
$3\times 10^{-2}$. From that I infer that non-perturbative contributions to 
$\alpha^{-1}$ should be $\lsim 0.006$. 

By ascribing a discrepancy in semileptonic $D$ decay data to OPE breaking
instanton effects, it has been argued~\cite{Shifman97} that instanton 
contributions could be as large as about 10\% in $R (m_\tau)$. Contributions
of this size are not excluded by the data, and would correspond to about
100\% of the PQCD corrections. Therefore, I will take 50\% of the PQCD 
correction to $I^{(3)}$ as a (conservative) uncertainty introduced by possible
OPE breaking effects (which are expected to be insignificant away from the real
axis). This yields an uncertainty of $\pm 0.006$ in $\alpha^{-1}$, 
incidentally identical to the bound obtained before. 

The SDR and UDR approaches are subject to the same size of non-perturbative 
effects, since we do not expect the relation between the on-shell and
\ms definitions of $\alpha (M_Z)$ to be afflicted by low energy effects. 
Indeed, the authors in Ref.~\cite{Davier98A} fitted different oscillating 
curves to the $R(s)$ data around $\mu_0$, and estimated the uncertainty to 
$\pm 0.002$ in $\alpha^{-1}$. This is of the same order of magnitude and 
smaller than the crude (and very conservative) estimate above. 

\section{Numerical Analysis}

In the numerical analysis I use $\hat\alpha_s(M_Z) = 0.120$ as a reference 
value. As mentioned earlier, no parametric error is included for 
$\hat\alpha_s$, which is regarded as a fit parameter. 

There is a variety of $c$~\cite{Dominguez94,Titard94} and 
$b$~\cite{Titard94,Davies94} running quark mass determinations. The 
uncertainties quoted by the various authors are of similar size, but being 
almost entirely theoretical, rather ad hoc. Therefore it does not seem 
justifiable to use a weighted average, which would yield\footnote{The 
various results are first converted to the scale invariant mass 
$\hat{m} (\hat{m})$, and averaged at the end.} 
$\hat{m}_c (\hat{m}_c) = 1.30 \pm 0.02$~GeV, i.e.\ a very small error. 
Instead, I determine the averages and uncertainties from the spread of the
results. This is a selfconsistent treatment, as it only needs the usual 
assumption of normal error distribution\footnote{It is also assumed that the 
determinations are approximately uncorrelated, as the various determinations
are very different. They range from quark potential methods and QCD sum rules 
to $D$ decays and lattice spectroscopy.}. The problem is then reduced to 
finding (posterior) information on a Gaussian distribution with (prior) unknown
mean and variance, when given $n$ data points (random drawings) $y_i$. It can 
be shown~\cite{Gelman95} that the marginal posterior distribution of the mean 
follows a Student-t distribution, which is centered at the sample mean, 
\be
   \bar{y} = {1\over n} \sum\limits_{i=1}^{n} y_i,
\ee
and has the standard deviation,
\be
   \sqrt{{1\over n(n-3)} \sum\limits_{i=1}^{n} (y_i - \bar{y})^2}.
\ee
Using this method I find,
\be
   \hat{m}_c (\hat{m}_c) = 1.31 \pm 0.07 \hbox{ GeV}, \\
\ee
\vspace{-28pt}
\be
   \hat{m}_b (\hat{m}_b) = 4.24 \pm 0.11 \hbox{ GeV}, 
\ee
which introduce uncertainties of $\pm 0.019$, and $\pm 0.002$ in 
$\hat\alpha^{-1} (M_Z)$, respectively. These are added linearly since they are
of similar origin. Notice, that the weighted and unweighted averages for
$\hat{m}_c$ are (fortuitously) in very good agreement with each other.

The ${\cal O} (\hat\alpha_s^2)$ term in Eq.~(\ref{int}) is clearly dominated
by the coefficient $-\beta_0^{(3)}\zeta(2)$. I will therefore use the analogous
coefficient in ${\cal O} (\hat\alpha_s^3)$, 
$-(\beta_1^{(3)} + 2 \beta_0^{(3)} r_2) \zeta(2) \approx -19$, as an estimate 
for the PQCD error in $I^{(3)}$, corresponding to $\pm 0.005$ in 
$\hat\alpha^{-1}$. I note that this large coefficient is not problematic for 
PQCD, as terms of this type can be resummed to all orders. The total 
theoretical uncertainty is $\pm 0.009$. Eq.~(\ref{SDR}) adds an experimental 
error of $\pm 0.015$. A variation of $\hat\alpha_s$ within $0.120\pm 0.005$ 
corresponds to $\hat\alpha^{-1} (M_Z) = 127.934^{-0.024}_{+0.020}$,
but this will not be included in the final error. 

The final result is
\be
\label{result}
\ba{r}\vspace{3pt}
   \hat\alpha^{-1} (m_\tau) = 133.513 \pm 0.015 \pm 0.009 \pm 0.019, \\
   \hat\alpha^{-1} (M_Z)    = 127.934 \pm 0.015 \pm 0.009 \pm 0.021,
\ea
\ee
where the errors are experimental, theoretical, and parametric, respectively. 
$\hat\alpha^{-1} (m_\tau)$ is to be compared with an earlier estimate
$\sim 133.29$~\cite{Marciano86} based on an ${\cal O} (\alpha)$ calculation
and $m_t = 45$ GeV. Using the ${\cal O} (\alpha\hat\alpha_s)$ conversion of
Ref.~\cite{Fanchiotti93} (with an ${\cal O} (\alpha\hat\alpha_s^2)$
improvement added), $\hat\alpha (M_Z)$ corresponds to
\be
   \Delta\bar\alpha_{\rm had}^{(5)} (M_Z) = 0.02779 \pm 0.00020.
\ee
Changing $\hat\alpha_s$ to 0.118 yields $0.02773$ in perfect agreement with 
$\Delta\bar\alpha_{\rm had}^{(5)} (M_Z) = 0.02774 \pm 0.00017$ from 
Ref.~\cite{Kuhn98}. 
Results for other values of $\mu \gsim \hat{m}_c$ can easily be obtained. 
A fit to all data using this approach yields for the Higgs mass,
\be
   M_H = 96_{-46}^{+76} \hbox{ GeV},
\ee 
compared to 
\be
   M_H = 69_{-43}^{+85} \hbox{ GeV},
\ee
from a fit to the same data set~\cite{Erler98}, 
but using $\bar\alpha (M_Z)$ from Ref.~\cite{Alemany98}.

\acknowledgements

It is a pleasure to thank Paul Langacker, Andr\'e Lukas and Dieter Zeppenfeld 
for discussions, and Andreas H\"ocker
for comments on Refs.~\cite{Davier98,Alemany98}. I greatly appreciate the 
hospitality of the Aspen Center for Physics where this work was initiated.

\end{document}